# A Linear-Transformer Hybrid for SNP-Based Genotype-to-Phenotype Prediction in Grapevine

**Yibin Wang[1], Murukarthick Jayakodi[2], Silvas Kirubakaran[3]. Ambika Chandra[2], *Azlan Zahid[1]**

[1] Department of Biological and Agricultural Engineering, Texas A&M AgriLife Research, Texas A&M University System, Dallas, TX, 75252, USA

[2] Department of Soil and Crop Sciences, Texas A&M AgriLife Research, Texas A&M University System, Dallas, TX, 75252, USA

[3]USDA-ARS, Grape Genetics Research Unit, 630 West North Street, Geneva, New York 14456, USA

*Corresponding author: azlan.zahid@tamu.edu

## Abstract

Robust genotype-to-phenotype (G2P) prediction is essential for accelerating breeding decisions and genetic gain. However, it remains challenging to measure complex traits under variable field conditions and across years. In this study, we propose a linear-Transformer approach, LiT-G2P (Linear-Transformer Genotype-to-Phenotype), an automated predictive framework that integrates additive genetic variance effects with Transformer-based nonlinear interactions using genome-wide single-nucleotide polymorphisms (SNPs) data. We evaluated LiT-G2P on a panel of diverse grape accessions, genotyped with SNP markers and measured for phenotypes across two consecutive years. Target phenotypic traits include leaf hair density and trichome density of grapevines. Across both single-year and cross-year testing scenarios, LiT-G2P consistently improves prediction performance compared with baseline models. For hair density, LiT-G2P achieves the lowest error in both single-year and cross-year evaluations, with RMSEs of 0.469 and 0.454, respectively, while maintaining strong tolerance accuracies of 79.2% and 74.6%, respectively. For trichome density, LiT-G2P also presents the best overall G2P performance. In addition, we extract model-prioritized SNPs from attention weights and apply genotype-stratified analysis to provide interpretable candidate marker for downstream validation. These results demonstrate that integrating stable additive effects with learned interaction patterns can enhance cross-year robustness and support practical SNP-based predictive modeling for genomic selection.

## 1. Introduction

Genotype-to-phenotype (G2P) relationships are fundamental for advancing agriculture by linking genetic variation to agronomically important traits (Fernandez-Pozo et al., 2015). Large-scale genotyping enables the discovery of genetic variants associated with agronomic traits (Bailey-Serres et al., 2019), disease resistance (Van Esse et al., 2020), and abiotic stress tolerance (Carvalho et al., 2021; Wang et al., 2003), providing a data-driven basis for selection decisions. Converting G2P associations into reliable predictions of complex phenotypes is essential for accelerating genetic gain in breeding programs, particularly for traits that are quantitative and influenced by environmental interactions. In G2P prediction, genomic inputs are typically encoded as genetic markers single-nucleotide polymorphisms (SNPs) which represent single-base DNA variants among individuals (Grinberg et al., 2020). Adopting such genotypic information for phenotype prediction can improve breeding efficiency, accelerate genetic gain, and support the development of resilient cultivars for sustainable agricultural production (Boggess et al., 2013).

Conventional crop breeding in perennial crops such as grapevine (*Vitis* spp.) relies on recurrent selection and evaluation across multiple environments to assemble favorable trait combinations (Eibach & Töpfer, 2015). Although improved cultivars have been produced, these approaches are inherently slow, resource intensive, and often limited in efficiently capturing quantitative traits (Callipo et al., 2025). Recent advances in phenotyping and next-generation sequencing transform plant breeding by enabling large-scale multidimensional data collection (Zhang et al., 2025). This breakthrough is driven by the development of phenotyping systems employing physical, chemical, and biological sensors, supported by emerging artificial intelligence (AI) technologies as well. Phenomics platforms now enable rapid, quantitative assessment of diverse responses, while long-read sequencing provides high-resolution genome structure and haplotype diversity (Demidchik et al., 2020). However, extensive genetic diversity and genomic complexity can complicate the dissection of trait architecture (Lappalainen et al., 2024). These challenges are further amplified by environmental effects, which can obscure causal signals and limit transferability. Such genetic and environmental heterogeneity jointly challenges the stability of genotype-phenotype associations, motivating robust predictive methods that can generalize G2P prediction for effective genomic selection and trait improvement in breeding.

AI and machine learning (ML) methods have advanced G2P modeling by integrating high dimensional genomic markers with complex and longitudinal phenotypic observations. Unlike conventional linear approaches, ML models can capture nonlinear genetic effects, epistatic interactions, and structured dependencies to improve explainable predictive performance for polygenic traits and enabling data-driven optimization of selection decisions (John et al., 2022; Novakovsky et al., 2023). A range of deep learning (DL) architectures including recurrent neural networks (RNNs) (Lugo & Hernández, 2021; Quang & Xie, 2016) and long short-term memory networks (LSTMs) (Tavakoli, 2019), and capsule networks (CapsNet) (Luo et al., 2023) have been explored to learn hierarchical feature representations from genomic data and to model temporal dynamics. Transformer-based architectures adopt self-attention to model long-range dependencies and have demonstrated strong utility across bioinformatics tasks including sequence representation learning, variant effect prediction, and multi-omics integration (Choi & Lee, 2023). By learning context-aware embeddings, attention mechanisms in Transformers and genome language models (gLMs) can prioritize informative genomic regions and flexibly aggregate signals across the genome, which fits sparse, heterogeneous, and high-dimensional breeding data (Consens et al., 2025). Transformer-based imputation methods employed sequential SNP data for human leukocyte antigen gene analysis (Tanaka et al., 2024). Recent advances in genomic foundation models further illustrate the potential of large-scale pretraining for transferable biological representations. DNABERT introduced bidirectional language modeling for DNA and learned generalizable sequence features from upstream and downstream nucleotide context (Ji et al., 2021). HyenaDNA expanded efficient long-context modeling to support single-nucleotide resolution with context windows reaching up to one million tokens, enabling representation learning over gene-scale and genome-scale spans (Nguyen et al., 2023). Similarly, the Nucleotide Transformer scaled model capacity to billions of parameters and leveraged training corpora spanning thousands of human genomes and hundreds of genomes across diverse species, improving cross-task transfer and robustness (Dalla-Torre et al., 2025). Despite these advances, predictive breeding frameworks that jointly utilize genomics, phenomics, and AI remain underdeveloped in agriculture crop plants. Moreover, while many genomic foundation models are trained on full DNA sequences, most applied breeding programs rely on cost-effective SNP genotyping (Mammadov et al., 2012), resulting in a gap between sequence-level representation learning and marker-based G2P prediction.

To address this gap, we propose an AI-empowered predictive breeding framework Linear-Transformer Genotype-to-Phenotype (LiT-G2P) that employs genome-wide SNP variants to support multi-trait improvement in grapevine. Building on our proposed LiT-G2P model, the predictor decomposes genetic effects into 1) additive main effects captured by a linear component and 2) nonlinear interactions modeled by a Transformer-based component, enabling simultaneous learning of interpretable marker contributions and dependencies across the genome. In this study, we focus on two related traits, hair density and trichome, formulating the prediction as a multi-trait learning problem with a shared Linear-Transformer backbone and a two-dimensional output layer. This design allows the model to exploit shared genetic architecture and phenotypic correlation between traits, effectively adopting shared information in the high-dimensional, low-sample regime ($N \ll p$). The linear main effect branch and Transformer interaction branch are trained across traits, while only the final output heads are trait specific. Such parameter sharing enables the backbone to learn genomic representations that are informative for both traits, acting as an additional form of regularization to improve robustness. Beyond grapevine, the proposed framework provides a practical template for integrating SNP-based G2P prediction into breeding pipelines for other crops where conventional selection is limited or inefficient. The main contributions of this study are presented as:

1. A novel SNP-based AI framework LiT-G2P that unifies linear additive main effects and nonlinear interaction Transformer-based modeling for grapevine G2P prediction, reflecting the genetic basis of complex traits.
2. A multi-trait formulation on cost-effective SNP markers that jointly predict target traits using a shared backbone and provides interpretability by selecting top-ranked SNPs and validating via genotype-stratified phenotype analysis.
3. A systematic evaluation against representative baselines on a grape SNP dataset to quantify the benefits of interaction modeling and multi-trait learning, demonstrating improved single-year and cross-year prediction for leaf hair density and trichome density score.

## 2. Materials and Methods

### 2.1 Data Acquisition

To establish feasibility, we assembled a diversity panel of 320 grape (*Vitis* spp.) accessions to capture broad genetic variation representative of breeding-relevant germplasm. Genotypic data were generated using a high-density SNP array from 2000 rhAmpSeq markers (Zou et al. 2020), producing 15,388 variants after standard quality control and filtering. In particular, SNPs and accessions were screened for overall call rate to remove unreliable genotyped markers and samples. Markers with low minor allele frequency of 0.05 were also excluded to reduce noise and improve statistical stability in the downstream modeling. We further filtered missing SNPs (6.45%) and removed variants with ambiguous allele coding or inconsistent genotype clustering. The resulting SNP set was curated, aligned across accessions, and formatted as a consistent genotype call/matrix (0/1/2) suitable for sequential modeling and cross-year evaluation. Complementary phenotypic measurements were collected over two consecutive growing seasons, 2023 and 2024, using BlackBird, an automated high-throughput phenotyping platform that enables repeatable field-based assessments at scale (Sapkota et al., 2025). In this study, we investigated two biologically related phenotypic traits, leaf hair density and trichome density. Figure 1 provides an overview of the phenotyping workflow and trait variability, illustrating BlackBird microscopic imaging of grapevine leaf discs for quantifying leaf hair and trichome density and representative leaf discs from diverse grape accessions highlighting broad phenotypic variation. The studied traits are relevant to plant protection and stress responses and can exhibit substantial genetic control while remaining sensitive to environmental conditions (Gago et al., 2016; Yang et al., 2024). Phenotypes were then evaluated and recorded to capture realistic variation arising from both genotype and environment. Each trait was scored on an ordinal scale from 0 to 3 in increments of 0.25, providing fine-grained resolution for quantitative modeling and enabling evaluation of predictive performance across years and traits. Figure 2 illustrates the phenotypic scoring criteria for the two target traits, showing representative examples of leaf hair density and trichome density.

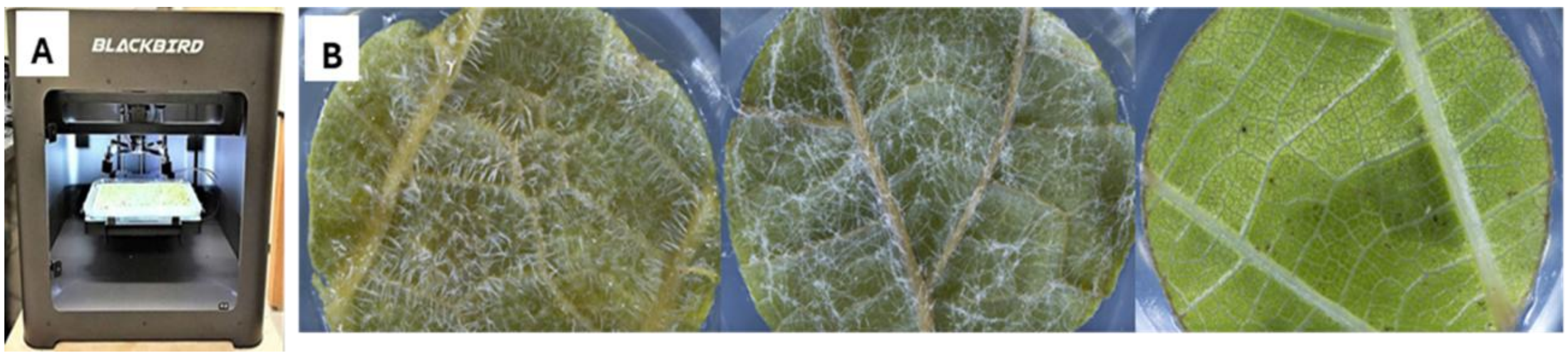

**Figure 1.** (A) BlackBird microscopic imaging of grapevine leaf discs to quantify leaf hair and trichome density; (B) Leaf discs of diverse grape accessions showing phenotypic variability for leaf hair and trichome density.

## 2.2 Data Preprocessing

For data preparation, we first processed Variant Call Format (VCF) genotype calls into SNP matrices and aligned them with phenotypic records using accession identifiers and year-specific observations. Genotypes were converted to alleles (0/1/2) to provide a consistent numerical representation for downstream modeling, where 0/0 refers to homozygous reference allele as 0, 0/1 or 1/0 refers to heterozygous allele as 1, and 1/1 refers to homozygous alternate allele as 2, respectively. This encoding yields an $N \times L$ matrix, where $N$ denotes the total number of accessions and $L$ is the number of SNP markers retained after quality control. Remaining missing genotype calls (6.45%) were imputed as major genotype to ensure a complete and consistent input matrix so that all accessions shared the same marker set and dimensionality. Phenotypic measurements were standardized to match the corresponding genotype entries, and records with missing trait scores were excluded from supervised training and evaluation. To model genome-wide dependencies beyond additive effects, we further implemented a Transformer-based architecture to capture both local and global linkage patterns across the genome. SNPs were grouped into non-overlapping patches of 64 SNPs with each patch embedded into a 128-dimensional latent space. Positional encodings were added to preserve chromosomal order across patches. This patch tokenization allows the proposed LiT-G2P to learn hierarchical SNP patterns while maintaining tractable attention computation over genomic inputs.

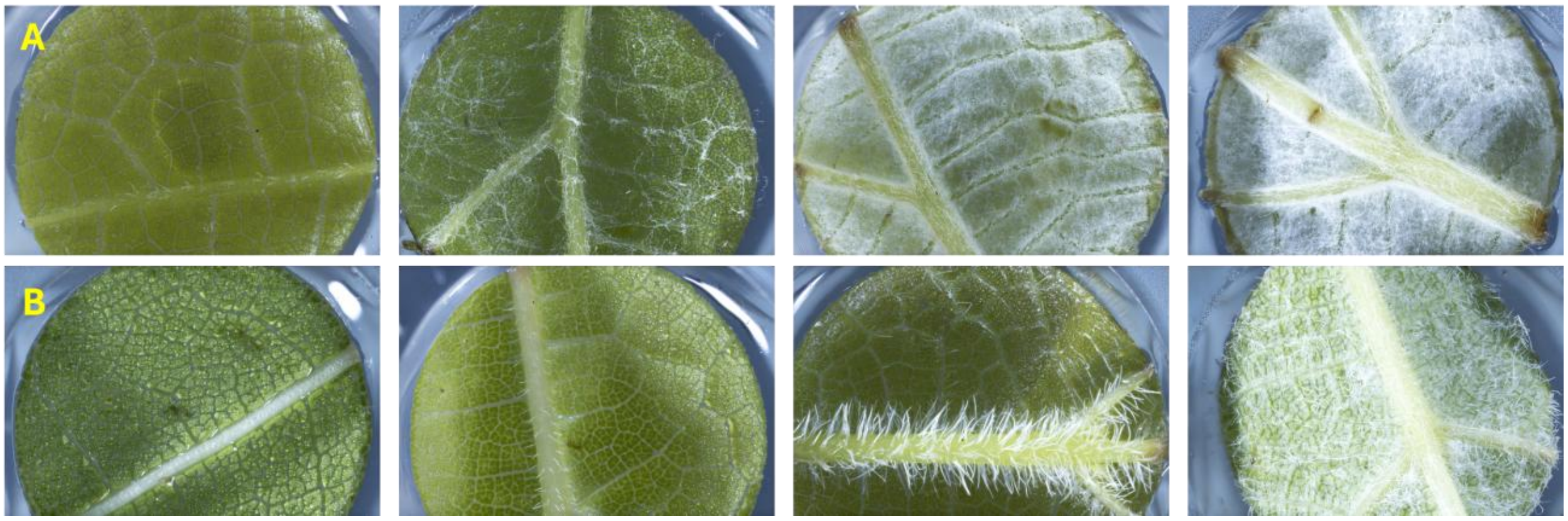

**Figure 2.** Phenotypic variability observed for leaf hair (A) and trichomes (B) densities were scored as 0, 1, 2, and 3 from left to right.

## 2.3 Methodology

### 2.3.1 The Proposed LiT-G2P Predictor

Hereby, we detail the proposed LiT-G2P framework for phenotypic trait prediction of grape leaf hair and trichome density from SNP data. This hybrid architecture decomposes phenotype prediction into additive main effects and nonlinear interaction effects. Given an accession $i$, the model takes as input the SNP vector $\mathbf{x}_i \in \mathbb{R}^L$, where the full genotype dataset can be represented by the matrix $\mathbf{X} \in \mathbb{R}^{N \times L}$. $N$ denotes the number of accessions and $L$ denotes the number of SNP variants. Each entry $X_{i,j} \in \{0,1,2\}$ corresponds to the allele dosage at SNP marker $j$ for individual $i$. The output $\boldsymbol{Y}$ of the model is represented by the phenotype vector $\mathbf{y}_i$ as,

$$\mathbf{y}_i = \begin{bmatrix} \boldsymbol{y}_i^{hair} \\ \boldsymbol{y}_i^{trich} \end{bmatrix} \in \mathbb{R}^2 \tag{1}$$

The overall phenotype prediction $\hat{\mathbf{y}}_i$ is modeled as,

$$\hat{\mathbf{y}}_i = \mathbf{W}_{\text{main}} \mathbf{x}_i + f_{TF}(\mathbf{x}_i) \tag{2}$$

$\mathbf{W}_{\text{main}} \in \mathbb{R}^{2\times L}$ are the additive SNP effects that describe the independent contribution of each gene's alleles to a trait. $f_{TF}(\cdot)$ is the Transformer-based interaction that learns epistasis interaction where the effect of one gene is dependent on the presence of one or more other genes. These two complementary components provide a biologically grounded decomposition of genetic architecture by combining interpretable main effects with higher-order interactions to better reflect how complex traits arise from genomic contributions. Specifically, the linear main effect branch provides a direct additive mapping from SNP dosages to trait values by,

$$\hat{\mathbf{y}}_i^{main} = \mathbf{W}_0\mathbf{x}_i \tag{3}$$

This branch models independent marker contributions and serves as a strong, interpretable baseline within the hybrid model. In the Transformer interaction branch, each SNP genotype value $X_{i,j} \in \{0,1,2\}$ is mapped into a continuous embedding as,

$$\mathbf{e}_{i,j} = \text{Embed}(\mathbf{X}_{i,j}) \in \mathbb{R}^D \tag{4}$$

The embeddings are stacked across all variants yielding,

$$\mathbf{E}_i = [\mathbf{e}_{i,1}, \mathbf{e}_{i,2}, \dots, \mathbf{e}_{i,L}] \in \mathbb{R}^{D\times L} \tag{5}$$

where $D$ refers to the embedding hidden dimension. To reduce the sequence length and learn local SNP patterns, variants are grouped into non-overlapping patches of size $P$. A depthwise 1D convolution produces patch tokens by,

$$\mathbf{Z}_i = \left[\mathbf{z}_{i,1}, \mathbf{z}_{i,2}, \dots, \mathbf{z}_{i,L'}\right] \in \mathbb{R}^{D\times L'} \tag{6}$$

where $L' = L/P$ is the number of patches given a patch size $P$. To preserve marker order across patches, sinusoidal positional embeddings $\mathbf{p}_q$ are applied by,

$$\tilde{\mathbf{z}}_{i,q} = \mathbf{z}_{i,q} + \mathbf{p}_q \tag{7}$$

where $\mathbf{p}_q$ is the sinusoidal positional embedding for token $q$ in $L'$. A standard Transformer encoder stack then models long-range dependencies among patches,

$$\mathbf{h}_{i,q} = \text{TransformerBlock}(\tilde{\mathbf{z}}_{i,1}, \tilde{\mathbf{z}}_{i,2}, \dots, \tilde{\mathbf{z}}_{i,L'}) \tag{8}$$

Patch-wise hidden states are aggregated into a single sequence representation with mean pooling as,

$$\bar{\mathbf{h}}_i = \frac{1}{L'}\sum_{q=1}^{L'} \mathbf{h}_{i,q} \in \mathbb{R}^D \tag{9}$$

The pooled representation is mapped to trait prediction using a multilayer perceptron (MLP) head by,

$$\hat{\mathbf{y}}_i^{inter} = \mathbf{W}_2\sigma\left(\mathbf{W}_1\bar{\mathbf{h}}_i(\mathbf{x}_i) + \mathbf{b}_1\right) + \mathbf{b}_2 \tag{10}$$

where $\mathbf{W}_1$ and $\mathbf{W}_2$ are trainable weight matrices, $\mathbf{b}_1$ and $\mathbf{b}_2$ are biases, and $\sigma(\cdot)$ is a pointwise nonlinearity. The resulting $\hat{\mathbf{y}}_i^{inter}$ captures nonlinear interaction-driven genetic effects. Therefore, the overall prediction can be represented as,

$$\hat{\mathbf{y}}_i = \mathbf{W}_0\mathbf{x}_i + \mathbf{W}_2\sigma\left(\mathbf{W}_1\bar{\mathbf{h}}_i(\mathbf{x}_i) + \mathbf{b}_1\right) + \mathbf{b}_2 \tag{11}$$

LiT-G2P performs joint prediction of hair density and trichome score with a shared backbone consisting of linear and Transformer branches. In implementation, the model outputs a two-dimensional vector $\mathbf{y}_i \in \mathbb{R}^2$ through a shared MLP head with two output units, allowing joint learning across traits and improving robustness $N \ll p$. Model parameters are optimized using mean squared error (MSE) over the two traits by the loss $\mathcal{L}$ as,

$$\mathcal{L} = \frac{1}{N}\sum_{i=1}^{N}\|\hat{\mathbf{y}}_i - \mathbf{y}_i\|_2^2 = \frac{1}{N}\sum_{i=1}^{N}[\left(\hat{y}_i^{hair} - y_i^{hair}\right)^2 + \left(\hat{y}_i^{trich} - y_i^{trich}\right)^2] \quad (12)$$

Regarding the model configuration and regularization, the encoder stack consists of 2 Transformer layers with 4 self-attention heads per layer, and a feedforward network of 256 hidden units. This design provides sufficient hierarchical abstraction of local-to-global SNP patterns while maintaining computational efficiency and stable convergence during training. The encoder employs multi-head self-attention to model dependencies among SNPs. Thus, the network can weigh both short-range and long-range interactions. Regularization was applied with a dropout rate of 0.2 and layer normalization. The final sequence representation was aggregated using a class token followed by a linear projection layer to predict trait values. Early stopping based on validation loss was employed to reduce overfitting. The end-to-end workflow of LiT-G2P is outlined in Algorithm 1.

**Algorithm 1** The proposed LiT-G2P predictor.

**Input:** Genotype matrix **X** with entries $X_{i,j}$; Phenotype matrix **Y** with vector $\mathbf{y}_i$; Patch size $P$; Embedding dimension $D$; Number of accessions $N$; Number of SNP variants $L$.
**Parameters:** $\mathbf{W}_0$; Embed $(\cdot)$; Positional embeddings $\mathbf{p}_q$; Transformer encoder $\mathrm{TF}(\cdot)$; MLP head $(\mathbf{W}_1, \mathbf{W}_2, \mathbf{b}_1, \mathbf{b}_2)$
**start**

- $L' \leftarrow L/P$
- **for** $i = 1$ to $N$ **do**
    - $\mathbf{x}_i \leftarrow \mathbf{X}_{i,:}^{\top} \in \mathbb{R}^L$
    - Main effects: $\hat{\mathbf{y}}_i^{main} = \mathbf{W}_0\mathbf{x}_i$
    - **for** $j = 1$ to $L$ **do**
        - $\mathbf{e}_{i,j} = \mathrm{Embed}(\mathbf{X}_{i,j})$
        - **end for**
    - $\mathbf{E}_i = [\mathbf{e}_{i,1}, \mathbf{e}_{i,2}, \dots, \mathbf{e}_{i,L}]$
    - $\mathbf{Z}_i = [\mathbf{z}_{i,1}, \mathbf{z}_{i,2}, \dots, \mathbf{z}_{i,L'}]$
    - **for** $q = 1$ to $L'$ **do**
        - $\tilde{\mathbf{z}}_{i,q} = \mathbf{z}_{i,q} + \mathbf{p}_q$
        - **end for**
    - $\mathbf{h}_{i,q} = \mathrm{TF}(\tilde{\mathbf{z}}_{i,1}, \tilde{\mathbf{z}}_{i,2}, \dots, \tilde{\mathbf{z}}_{i,L'})$
    - Pooling: $\bar{\mathbf{h}}_i = \frac{1}{L'}\sum_{q=1}^{L'}\mathbf{h}_{i,q}$
    - $\hat{\mathbf{y}}_i^{inter} = \mathbf{W}_2\sigma\left(\mathbf{W}_1\bar{\mathbf{h}}_i(\mathbf{x}_i) + \mathbf{b}_1\right) + \mathbf{b}_2$
    - **output** Final prediction: $\hat{\mathbf{y}}_i$
- **end for**
- $\mathbf{Y} \leftarrow [\hat{\mathbf{y}}_1^{\top}, \hat{\mathbf{y}}_2^{\top}, \dots, \hat{\mathbf{y}}_N^{\top}]^{\top}$

**end**

2.3.2 Performance Evaluation

Initial experiments with lightweight LiT-G2P demonstrated feasibility but were limited by small sample size and underfitting. By computing allele-specific embeddings directly from the VCF data, we generated biologically meaningful feature representations without requiring whole-genome sequences. These embeddings were aggregated across the genome and coupled with regularized regression to provide robust trait prediction under limited sample sizes. We trained the LiT-G2P with MSE loss for continuous traits and monitored performance with RMSE, MAE, and MAPE. In addition to conventional performance metrics, we evaluated categorical accuracy within error thresholds of 0.5 units. This bin-based approach reflects practical breeding utility, where predictions that fall within half a unit of the observed phenotype are still considered actionable for selection. Consequently, we assessed model performance not only in terms of statistical fit but also in terms of decision-making reliability for downstream breeding applications.

We evaluate the prediction performance of the proposed method against the baseline G2P prediction methods including linear ridge regression, XGBoost, and baseline Transformer without modeling main effects. Two evaluation protocols were conducted: 1) a single-year evaluation of 2024 data with an 80%/20% training/test stratified partition for model selection, and 2) a cross-year validation while keeping the same training data in 2024 and testing in 2023. Relevant performance metrics are calculated in Equation (13) – (16) as,

$$\mathrm{RMSE}(\hat{\mathbf{y}}_i, \mathbf{y}_i) = \sqrt{\frac{1}{N}\sum_{i=1}^{N}(\mathbf{y}_i - \hat{\mathbf{y}}_i)^2} \tag{13}$$

$$\mathrm{MAE}(\hat{\mathbf{y}}_i, \mathbf{y}_i) = \frac{1}{N}\sum_{i=1}^{N}|\mathbf{y}_i - \hat{\mathbf{y}}_i| \tag{14}$$

$$\mathrm{MAPE}(\hat{\mathbf{y}}_i, \mathbf{y}_i) = \frac{1}{N}\sum_{i=1}^{N}\left|\frac{\mathbf{y}_i - \hat{\mathbf{y}}_i}{\mathbf{y}_i + \epsilon}\right| \tag{15}$$

$$\mathrm{ACC}_{\pm 0.5}(\hat{\mathbf{y}}_i, \mathbf{y}_i) = \frac{1}{N}\sum_{i=1}^{N}\mathbf{1}(|\mathbf{y}_i - \hat{\mathbf{y}}_i| \le 0.5) \tag{16}$$

where $\mathbf{y}_i$ and $\hat{\mathbf{y}}_i$ refer to the actual and predicted plant phenotypic trait vectors, respectively. $\bar{\mathbf{y}}_i$ is the sample mean, and $\epsilon$ refers to the constant to avoid division by zero in MAPE. In addition to plant-wise prediction, we also evaluate the model at the population level by comparing the predicted and observed trait-score distributions across accessions, assessing whether the model preserves overall phenotypic variability and ranking patterns.

To improve interpretability, we extracted the Top-$k$ informative SNPs prioritized by the LiT-G2P attention mechanism. Because the Transformer operates on patch tokens, we first aggregated attention weights across heads and layers to obtain a single importance score per token. Specifically, let $\mathbf{A}^{(m,h)}$ denote the self-attention matrix from layer $m$ and head $h$. A patch-level relevance score is computed by averaging attention across heads and layers to define the importance score $s_k$ as,

$$s_k = \frac{1}{MH}\sum_{m=1}^{M}\sum_{h=1}^{H}\mathbf{A}_{q,k}^{(m,h)} \tag{17}$$

This SNP importance can therefore be quantified by aggregating patch-level attention weights across layers and heads, mapping each SNP to its corresponding patch score. In evaluation, we rank and report the top-$k$ informative SNPs according to the main-effect weights for the target traits, highlighting markers with the strongest estimated genetic effects.

## 3 Results and Discussion

Performance of single-year and cross-year prediction is shown in Table 1 – 2 for leaf hair density and trichome density across different model families. Across both traits, single-year evaluation yields consistently better performance than cross-year testing, indicating that year-to-year variation introduces a meaningful distribution shift in the G2P mapping. This effect is especially evident for the classical baselines that linear ridge and XGBoost show notable degradations in cross-year accuracy, e.g., hair density $ACC_{\pm 0.5}$ drops to 50 to 52%, suggesting limited ability to generalize when phenotypic expression changes across seasons. In contrast, Transformer-based models exhibit improved resilience under cross-year evaluation, maintaining substantially lower error and higher tolerance accuracy, consistent with their capacity to learn distributed, nonlinear dependencies among SNP markers.

For hair density prediction according to Table 1, LiT-G2P achieves the strongest overall performance in both scenarios, with the lowest RMSE and MAPE in single-year testing of 0.469 and 26.4%, respectively. The proposed method also remains the least cross-year degradation with 0.454 cross-year RMSE and 74.6% $ACC_{\pm 0.5}$. Notably, the baseline Transformer improves over ridge and XGBoost, yet LiT-G2P provides an additional improvement, supporting the benefit of decomposing genetic effects into additive main effects plus Transformer-modeled interactions. While the Transformer attains slightly higher accuracy than LiT-G2P in the single-year case, LiT-G2P reduces continuous prediction error more strongly, i.e., lower RMSE, MAE, and MAPE, indicating less deviations and better calibration even when small threshold boundary shifts affect the categorical tolerance metric.

**Table 1.** Comparison of prediction performance for leaf hair density under single-year and cross-year.

| Testing Scenario | Method | RMSE | MAE | MAPE | $ACC_{\pm 0.5}$ |
|---|---|---|---|---|---|
| **Single-Year** | Linear Ridge | 0.758 | 0.484 | 47.9% | 68.8% |
| | XGBoost | 0.722 | 0.469 | 44.5% | 67.2% |
| | Transformer | 0.608 | 0.386 | 30.3% | 81.3% |
| | **LiT-G2P (ours)** | **0.469** | **0.363** | **26.4%** | **79.2%** |
| **Cross-Year** | Linear Ridge | 0.792 | 0.594 | 56.2% | 49.8% |
| | XGBoost | 0.753 | 0.581 | 52.8% | 52.4% |
| | Transformer | 0.647 | 0.516 | 37.1% | 69.5% |
| | **LiT-G2P (ours)** | **0.454** | **0.392** | **31.5%** | **74.6%** |

Trichome density prediction performance is uniformly lower than for hair density as demonstrated in Table 2. This indicates the increasing difficulty of this trait, potentially due to stronger environmental sensitivity, measurement variability, or a more complex genetic architecture. Nevertheless, LiT-G2P presents the best prediction performance under both evaluation regimes, outperforming the baseline Transformer and the classical methods in RMSE, MAE, and accuracy. Furthermore, cross-year robustness is improved with RMSE of 0.760 and $ACC_{\pm 0.5}$ of 57.8%. Collectively, these results suggest that the proposed LiT-G2P framework provides a practical advantage for breeding-oriented prediction, particularly when generalization across years is required, by combining stable additive signals with flexible modeling of dependencies in high dimensional SNP data.

**Table 2.** Comparison of prediction performance for trichome density under single-year and cross-year.

| Testing Scenario | Method | RMSE | MAE | MAPE | $ACC_{\pm 0.5}$ |
|---|---|---|---|---|---|
| **Single-Year** | Linear Ridge | 0.925 | 0.754 | 55.8% | 50.0% |
| | XGBoost | 0.741 | 0.575 | 49.2% | 57.9% |
| | Transformer | 0.713 | 0.579 | 45.9% | 59.4% |
| | **LiT-G2P (ours)** | **0.685** | **0.554** | **44.2%** | **62.6%** |
| **Cross-Year** | Linear Ridge | 0.941 | 0.763 | 58.9% | 45.3% |
| | XGBoost | 0.912 | 0.736 | 54.2% | 50.5% |
| | Transformer | 0.806 | 0.597 | 48.6% | 51.6% |
| | **LiT-G2P (ours)** | **0.760** | **0.562** | **45.3%** | **57.8%** |

The impact of cross-year evaluation on G2P prediction is summarized in Figure 3 by plotting the change in performance relative to single-year testing (Δ = cross-year − single-year) for two target traits. Positive values for RMSE/MAE/MAPE indicate increased prediction error under cross-year transfer, while negative values for accuracy indicate reduced classification-style agreement, jointly visualizing the robustness gap around a zero baseline. Across both traits, conventional baselines show larger error inflation and accuracy degradation. Transformer-based models achieve improved cross-year stability, with LiT-G2P consistently producing smaller error increases and accuracy drops, suggesting that representation learning better captures year-stable genomic signals and mitigates distribution shift. The presented layout provides an overall comparison of cross-year generalization and highlights that the proposed LiT-G2P most effectively preserves predictive performance across years.

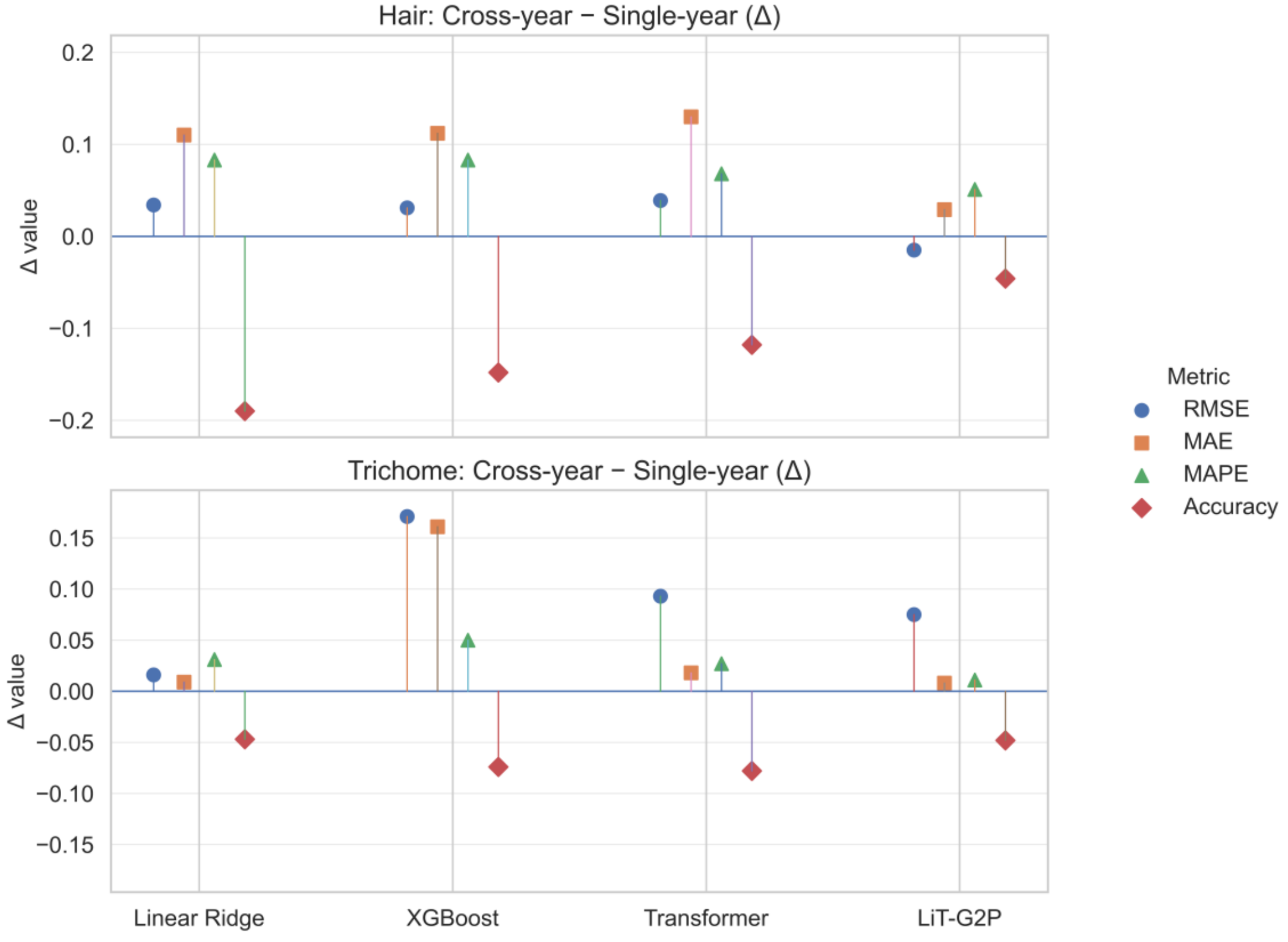

**Figure 3.** Diverging generalization shift from single-year to cross-year prediction: error increases (Δ>0) and accuracy decreases (Δ<0) across models and traits.

A key limitation revealed in the testing is that predictive performance degrades under cross-year evaluation, indicating sensitivity to environmental shifts in phenotypic expression and measurement conditions. Although LiT-G2P and the baseline Transformer generalize better than linear ridge and tree models, both still show increased error and reduced accuracy when models trained on one season are applied to another, consistent with genotype-by-environment effects and potential differences in field microclimate, management, or imaging conditions across years. This limitation is particularly observed for trichome density, where overall errors remain higher, and tolerance accuracy is lower than hair density in both single-year and cross-year settings.

**Table 3.** Top 5 informative SNPs prioritized by the LiT-G2P attention mechanism.

| Index | #CHROM | POS | ID | REF | ALT | INFO | FORMAT |
|---|---|---|---|---|---|---|---|
| 6729 | 8 | 21324057 | SPN8_21324057 | G | A | AC=127; AN=638 | GT:DS |
| 6617 | 8 | 18542656 | SPN8_18542656 | A | G | AC=109; AN=564 | GT:DS |
| 4943 | 7 | 1394640 | SPN7_1394640 | G | A | AC=105; AN=634 | GT:DS |
| 6616 | 8 | 18542650 | SPN8_18542650 | T | C | AC=108; AN=564 | GT:DS |
| 955 | 1 | 26601686 | SPN1_26601686 | A | G | AC=31; AN=624 | GT:DS |

Table 3 reports the top 5 SNPs most prioritized by the LiT-G2P attention mechanism, providing an interpretable set of candidate makers that the model repeatedly relies on when predicting the target traits. Notably, the top-ranked markers are concentrated on chromosomes #7 and #8 with three out of five on chromosome #8, suggesting that a subset of genomic regions may carry disproportionately informative signals for leaf hair and trichome variation in this diversity panel. The presence of two nearby variants at positions 18,542,650 and 18,542,656 on chromosome #8 further indicates a localized block or tightly linked region that the model considers important, consistent with the expectation that clustered SNPs can tag the same underlying causal variants. The allele summaries show that these prioritized SNPs span moderate to lower allele frequencies, implying that the model is not solely driven by common variants but can also utilize less frequent polymorphisms when they contribute predictive signal. While attention-based prioritization does not establish causality by itself, these SNPs serve as a focused starting point for downstream validation, such as testing genotype-stratified phenotype differences, mapping to nearby genes, or conducting targeted ablation analysis to quantify the change in prediction error.

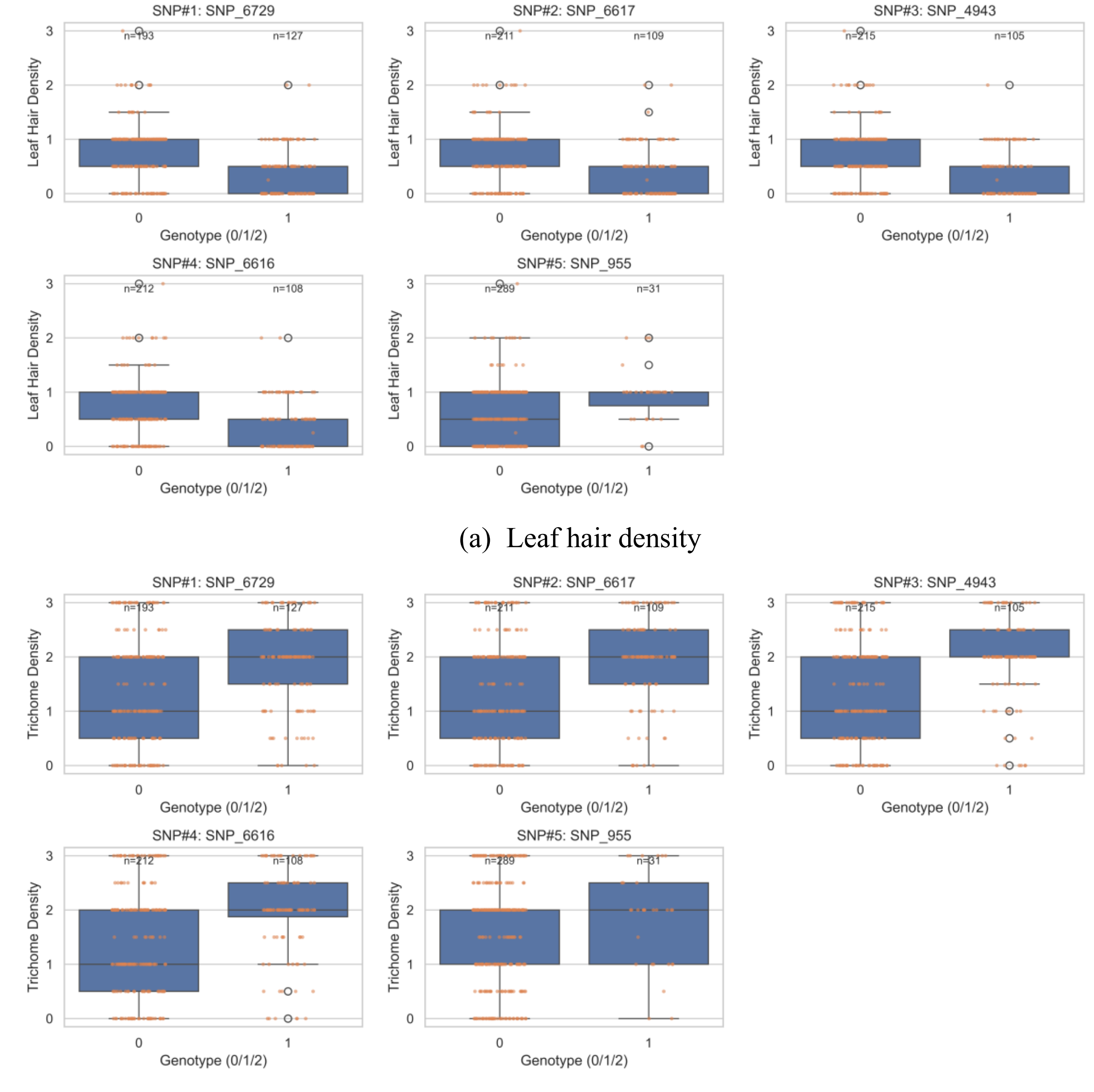

(a) Leaf hair density

(b) Trichome density

**Figure 4.** Genotype-stratified (a) leaf hair density and (b) trichome density for the top five SNPs.

The genotype-stratified boxplots shown in Figure 4 provide an interpretable check of the top-ranked SNPs by showing how trait values vary across genotype classes (0/1/2). Strongly associated SNPs are expected to stratify the phenotype by genotype class, producing clearly separated distributions including shifted medians and reduced overlap between genotype groups. According to the plot, it is observed that leaf hair density shows clearer genotype-dependent separation than trichome density across the top SNPs. This stratification is consistent with the better predictive performance observed for leaf hair with higher accuracy, suggesting that leaf hair is driven by more readily learnable and stable genetic signal in this panel. Specifically, leaf hair density shows an overall trend for four out of the five SNPs, where genotype-1 is associated with lower hair density relative to genotype-0. The fifth SNP at position 26,601,686 displays a weaker reversed pattern, highlighting potential trait-dependent effects. For trichome density, all five SNPs exhibit a consistent shift in distribution between genotype groups, with the genotype-1 group generally showing higher trichome density than genotype-0, suggesting a coherent allele-dose effect for the alternate allele across markers. Overall, the repeated genotype-dependent separation supports that the selected SNPs capture meaningful genetic signal aligned with phenotypic variation in the panel.

LiT-G2P offers a practical and interpretable solution for improving SNP-based prediction in grapevine. Integrating additive genetic effects with Transformer-modeled interactions improved both the single-year accuracy and cross-year robustness of the model while providing biologically meaningful marker interpretation through attention-based SNP prioritization. These capabilities make LiT-G2P well-suited for breeding pipelines, where earlier identification of elite accessions, reduced phenotyping burden, and greater stability across variable environments are critical for accelerating selection decisions. Despite these advantages, several limitations warrant further investigation. Although the overall SNP missingness was low, we imputed remaining missing calls using the major genotype predominantly 0 to obtain a fixed-dimensional SNP matrix for all accessions. This pragmatic strategy may introduce bias if missingness is non-random or if rare-allele patterns are affected. Future work will evaluate more principled imputation schemes and explicitly assess the sensitivity of predictive performance to alternative missing-data treatments. Imbalanced genotype classes among high-importance SNPs may restrict effect estimation, emphasizing the need for larger and more diverse panels. In addition, the current analysis focuses on two years and two traits. Extending to additional environments, years, management, and broader trait suites will better characterize robustness under various distribution shifts. Future work should explore additional modeling strategies that incorporate genotype-by-environment interactions to better capture environment-dependent genetic effects. Pretraining the Transformer encoder on large multi-crop genotype datasets also represents a promising direction toward developing a general-purpose, foundation-style SNP model that can be efficiently adapted across crops and traits.

## 4 Conclusion

This study presents LiT-G2P, a hybrid G2P predictor that decomposes genetic architecture into additive main effects and Transformer-modeled nonlinear interactions, enabling accurate multi-trait prediction from genome-wide SNP data in grapevine. Overall, LiT-G2P demonstrates that integrating additive and interaction modeling within an attention-based architecture can improve robustness and utility of SNP-based G2P prediction. A diversity panel with high-dimensional genotypes and two years of field phenotyping is investigated. LiT-G2P consistently outperforms linear ridge, XGBoost, and a Transformer baseline for both leaf hair density and trichome density. The improvement is observed in cross-year evaluation, where prediction robustness is critical for deployment-oriented breeding decisions. LiT-G2P maintains higher tolerance accuracy and smaller performance degradation than competing methods, suggesting improved generalization under year-to-year shifts. Beyond predictive accuracy, LiT-G2P provides a practical pathway toward interpretable AI-assisted breeding and selection. The linear branch yields trait-specific additive marker effects, while attention-derived importance scores prioritize genomic regions and SNPs that contribute most to model predictions. Genotype-stratified analysis of top SNPs reveals clear phenotype separation, supporting that the learned signals are biologically meaningful and consistent with the observed predictive performance. This provides targeted candidates for biological follow-up. Nevertheless, cross-year performance remains limited by environmental variability and phenotyping noise, which highlights the demand for larger multi-year datasets and explicit modeling of genotype-by-environment effects. Future work will extend LiT-G2P to multi-environment training, domain-shift regularization, and validate prioritized SNPs using genotype-stratified analysis on external populations.

**Acknowledgments**

The authors thank Anna Underhill of the Lance Cadle-Davidson USDA-ARS laboratory for assistance with leaf sample collection and Blackbird robot measurements. The authors also thank Erin Galarneau for ensuring that the necessary resources were available for these experiments.

**Authors' Contributions**

**Yibin Wang:** Conceptualization, Data curation, Methodology, Investigation, Validation, Writing – Original draft; **Murukarthick Jayakodi:** Conceptualization, Data curation, Writing – reviewing & editing; **Silvas Kirubakaran:** Conceptualization, Data curation, Writing – reviewing & editing; **Ambika Chandra:** Conceptualization, Writing – reviewing & editing; **Azlan Zahid:** Conceptualization, Methodology, Investigation, Writing – reviewing & editing, Resources, Supervision, Funding acquisition.

**Funding**

This research is partially supported by the United States Department of Agriculture (USDA)'s National Institute of Food and Agriculture (NIFA) Research Capacity Fund Hatch Program: TEX09954 (Accession No. 7002248) and Research Capacity Fund Multistate Hatch Program: TEX0-1-9916 (Accession No. 7008389). Leaf phenotypic data collection was supported by the USDA-ARS project Genetic Improvement of Grape Quality and Adaptation to Diseases and Abiotic Stress (Project No. 8060-21220-008-000-D). Any opinions, findings, conclusions, or recommendations expressed in this publication are those of the authors and should not be construed to represent any official USDA or U.S. Government determination or policy.

**Data Availability**

The data will be made available on request.

**References**

Bailey-Serres, J., Parker, J. E., Ainsworth, E. A., Oldroyd, G. E. D., & Schroeder, J. I. (2019). Genetic strategies for improving crop yields. *Nature*, *575*(7781), 109–118. https://doi.org/10.1038/s41586-019-1679-0

Boggess, M. V., Lippolis, J. D., Hurkman, W. J., Fagerquist, C. K., Briggs, S. P., Gomes, A. V., Righetti, P. G., & Bala, K. (2013). The need for agriculture phenotyping: "Moving from genotype to phenotype." *Journal of Proteomics*, *93*, 20–39. https://doi.org/10.1016/j.jprot.2013.03.021

Callipo, P., Schmidt, M., Strack, T., Robinson, H., Vasudevan, A., & Voss-Fels, K. P. (2025). Harnessing clonal diversity in grapevine: From genomic insights to modern breeding applications. *Theoretical and Applied Genetics*, *138*(8), 196. https://doi.org/10.1007/s00122-025-04986-w

Carvalho, L. C., Gonçalves, E. F., Marques Da Silva, J., & Costa, J. M. (2021). Potential Phenotyping Methodologies to Assess Inter- and Intravarietal Variability and to Select Grapevine Genotypes Tolerant to Abiotic Stress. *Frontiers in Plant Science*, *12*, 718202. https://doi.org/10.3389/fpls.2021.718202

Choi, S. R., & Lee, M. (2023). Transformer Architecture and Attention Mechanisms in Genome Data Analysis: A Comprehensive Review. *Biology*, *12*(7), 1033. https://doi.org/10.3390/biology12071033

Consens, M. E., Dufault, C., Wainberg, M., Forster, D., Karimzadeh, M., Goodarzi, H., Theis, F. J., Moses, A., & Wang, B. (2025). Transformers and genome language models. *Nature Machine Intelligence*, *7*(3), 346–362. https://doi.org/10.1038/s42256-025-01007-9

Dalla-Torre, H., Gonzalez, L., Mendoza-Revilla, J., Lopez Carranza, N., Grzywaczewski, A. H., Oteri, F., Dallago, C., Trop, E., De Almeida, B. P., Sirelkhatim, H., Richard, G., Skwark, M., Beguir, K., Lopez, M., & Pierrot, T. (2025). Nucleotide Transformer: Building and evaluating robust foundation models for human genomics. *Nature Methods*, *22*(2), 287–297. https://doi.org/10.1038/s41592-024-02523-z

Demidchik, V. V., Shashko, A. Y., Bandarenka, U. Y., Smolikova, G. N., Przhevalskaya, D. A., Charnysh, M. A., Pozhvanov, G. A., Barkosvkyi, A. V., Smolich, I. I., Sokolik, A. I., Yu, M., & Medvedev, S. S. (2020). Plant Phenomics: Fundamental Bases, Software and Hardware Platforms, and Machine Learning. *Russian Journal of Plant Physiology*, *67*(3), 397–412. https://doi.org/10.1134/S1021443720030061

Eibach, R., & Töpfer, R. (2015). Traditional grapevine breeding techniques. In *Grapevine Breeding Programs for the Wine Industry* (pp. 3–22). Elsevier. https://doi.org/10.1016/B978-1-78242-075-0.00001-6

Fernandez-Pozo, N., Menda, N., Edwards, J. D., Saha, S., Tecle, I. Y., Strickler, S. R., Bombarely, A., Fisher-York, T., Pujar, A., Foerster, H., Yan, A., & Mueller, L. A. (2015). The Sol Genomics Network (SGN)—From genotype to phenotype to breeding. *Nucleic Acids Research*, *43*(D1), D1036–D1041. https://doi.org/10.1093/nar/gku1195

Gago, P., Conéjéro, G., Martínez, M. C., Boso, S., This, P., & Verdeil, J.-L. (2016). Microanatomy of leaf trichomes: Opportunities for improved ampelographic discrimination of grapevine ( *Vitis vinifera* L.) cultivars: Grapevine leaf trichomes. *Australian Journal of Grape and Wine Research*, *22*(3), 494–503. https://doi.org/10.1111/ajgw.12226

Grinberg, N. F., Orhobor, O. I., & King, R. D. (2020). An evaluation of machine-learning for predicting phenotype: Studies in yeast, rice, and wheat. *Machine Learning*, *109*(2), 251–277. https://doi.org/10.1007/s10994-019-05848-5

Ji, Y., Zhou, Z., Liu, H., & Davuluri, R. V. (2021). DNABERT: Pre-trained Bidirectional Encoder Representations from Transformers model for DNA-language in genome. *Bioinformatics*, *37*(15), 2112–2120. https://doi.org/10.1093/bioinformatics/btab083

John, M., Haselbeck, F., Dass, R., Malisi, C., Ricca, P., Dreischer, C., Schultheiss, S. J., & Grimm, D. G. (2022). A comparison of classical and machine learning-based phenotype prediction methods on simulated data and three plant species. *Frontiers in Plant Science*, *13*, 932512. https://doi.org/10.3389/fpls.2022.932512

Lappalainen, T., Li, Y. I., Ramachandran, S., & Gusev, A. (2024). Genetic and molecular architecture of complex traits. *Cell*, *187*(5), 1059–1075. https://doi.org/10.1016/j.cell.2024.01.023

Lugo, L., & Hernández, E. B.-. (2021). A Recurrent Neural Network approach for whole genome bacteria identification. *Applied Artificial Intelligence*, *35*(9), 642–656. https://doi.org/10.1080/08839514.2021.1922842

Luo, X., Kang, X., & Schönhuth, A. (2023). Predicting the prevalence of complex genetic diseases from individual genotype profiles using capsule networks. *Nature Machine Intelligence*, *5*(2), 114–125. https://doi.org/10.1038/s42256-022-00604-2

Mammadov, J., Aggarwal, R., Buyyarapu, R., & Kumpatla, S. (2012). SNP Markers and Their Impact on Plant Breeding. *International Journal of Plant Genomics*, *2012*, 1–11. https://doi.org/10.1155/2012/728398

Nguyen, E., Poli, M., & Faizi, M. (2023). *HyenaDNA: Long-Range Genomic Sequence Modeling at Single Nucleotide Resolution*.

Novakovsky, G., Dexter, N., Libbrecht, M. W., Wasserman, W. W., & Mostafavi, S. (2023). Obtaining genetics insights from deep learning via explainable artificial intelligence. *Nature Reviews Genetics*, *24*(2), 125–137. https://doi.org/10.1038/s41576-022-00532-2

Quang, D., & Xie, X. (2016). DanQ: A hybrid convolutional and recurrent deep neural network for quantifying the function of DNA sequences. *Nucleic Acids Research*, *44*(11), e107–e107. https://doi.org/10.1093/nar/gkw226

Sapkota, S., Martinez, D., Underhill, A., Chen, L.-L., Gadoury, D., Cadle-Davidson, L., & Hwang, C.-F. (2025). A Device for Computer Vision Analysis of Fungal Features Outperforms Quantitative Manual Microscopy by Experts in Discerning a Host Resistance Locus. *Phytopathology*, *115*(10), 1357–1367. https://doi.org/10.1094/PHYTO-01-25-0033-R

Tanaka, K., Kato, K., Nonaka, N., & Seita, J. (2024). Efficient HLA imputation from sequential SNPs data by transformer. *Journal of Human Genetics*, *69*(10), 533–540. https://doi.org/10.1038/s10038-024-01278-x

Tavakoli, N. (2019). Modeling Genome Data Using Bidirectional LSTM. *2019 IEEE 43rd Annual Computer Software and Applications Conference (COMPSAC)*, 183–188. https://doi.org/10.1109/COMPSAC.2019.10204

Van Esse, H. P., Reuber, T. L., & Van Der Does, D. (2020). Genetic modification to improve disease resistance in crops. *New Phytologist*, *225*(1), 70–86. https://doi.org/10.1111/nph.15967

Wang, W., Vinocur, B., & Altman, A. (2003). Plant responses to drought, salinity and extreme temperatures: Towards genetic engineering for stress tolerance. *Planta*, *218*(1), 1–14. https://doi.org/10.1007/s00425-003-1105-5

Yang, B., Liu, J., Gu, Q., Xu, Z., Yao, X., Liang, J., Xu, M., Lu, J., & Fu, P. (2024). Identification of the LH2 Locus for Prostrate Hair Density in Grapevine. *Horticulturae*, *10*(12), 1309. https://doi.org/10.3390/horticulturae10121309

Zhang, Y., Huang, G., Zhao, Y., Lu, X., Wang, Y., Wang, C., Guo, X., & Zhao, C. (2025). Revolutionizing Crop Breeding: Next-Generation Artificial Intelligence and Big Data-Driven Intelligent Design. *Engineering*, *44*, 245–255. https://doi.org/10.1016/j.eng.2024.11.034

Zou, C., Karn, A., Reisch, B., Nguyen, A., Sun, Y., Bao, Y., ... & Cadle-Davidson, L. (2020). Haplotyping the Vitis collinear core genome with rhAmpSeq improves marker transferability in a diverse genus. Nature communications, 11(1), 413.